\documentclass{iau}
\usepackage{graphicx}
\usepackage{amsmath}
\usepackage{mathrsfs}
\usepackage{dblfloatfix}
\usepackage{color}
\usepackage{fixltx2e}
\usepackage{float}
\usepackage{placeins}
\usepackage{array, makecell} 
\usepackage{gensymb}
\usepackage{booktabs}
\usepackage{url}
\usepackage{subfloat}
\usepackage[version=3]{mhchem}
\usepackage{threeparttable}
\usepackage{multirow}
\usepackage{CJK}
\usepackage{longtable}
\usepackage{pdflscape}
\usepackage[export]{adjustbox}
\usepackage{subfig}
\usepackage{array}
\usepackage{bm}
\usepackage[singlelinecheck=false, labelfont=bf]{caption}
\usepackage{natbib,twoopt}

\newcommand{\kms}{{\hbox {\,km\,s$^{-1}$}}}
\newcommand{\arcsec}{{\hbox {${''}$}}}

\newcommand{\lsun}{{\hbox {$L_\odot$}}}
\newcommand{\msun}{{\hbox {$M_\odot$}}}

\label{key}

\def\ttco#1#2{{\hbox {${\mathrm{^{13}CO}}(#1\text{--}#2)$}}}
\def\hcn#1#2{{\hbox {${\mathrm{HCN}}(#1\text{--}#2)$}}}
\def\hcop#1#2{{\hbox {${\mathrm{HCO^+}}(#1\text{--}#2)$}}}
\newcommand*\aap{A\&A}

\newcommand*\apj{ApJ}
\newcommand*\apjl{ApJ}

\newcommand*\mnras{MNRAS}

\newcommand*\nat{Nature}

\title[5--12\,pc view of the compact obscured nucleus of IRAS\,17578-0400] %% give here short title %%
{5--12 pc resolution ALMA imaging of gas and dust in the obscured compact nucleus of IRAS\,17578-0400}

\author[Chentao Yang \& the CONfirm team]   %% give here short author list %%
{Chentao Yang$^{1}$                 ,     
Susanne Aalto$^{1}$                 , 
Sabine K\"onig$^{1}$                , \\
Santiago Del Palacio$^{1}$          , 
Mark Gorski$^{1}$                   , 
Sean Linden$^{2}$                   , \\
Sebastien Muller$^{1}$              , 
Kyoko Onishi$^{1}$                  , 
Mamiko Sato$^{1}$                   , 
Clare Wethers$^{1}$                 }

\affiliation{$^1$Department of Space, Earth and Environment, \\ Chalmers University of Technology, Onsala Space Observatory, 439 92 Onsala, Sweden \\ email: {\tt chentao.yang@chalmers.se} \\[\affilskip]
$^2$Department of Astronomy, University of Massachusetts at Amherst, \\ Amherst, MA 01003, USA}

\pubyear{2023}
\volume{378}  %% insert here IAU Symposium No.
\pagerange{xxx--xxx}
\jname{Black hole winds at all scales}
\editors{G. Bruni, M. Diaz-Trigo, S. Laha, and K. Fukumura}
\begin{document}

\maketitle

\begin{abstract}
We here present 0.02--0.04\arcsec\ resolution ALMA observation of the compact obscured nucleus (CON) of IRAS\,17578-0400. A dusty torus within the nucleus, approximately 4\,pc in radius, has been uncovered, exhibiting a usually flat spectral index at ALMA band 3, likely due to the millimeter corona emission from the central supermassive black hole (SMBH). The dense gas disk, traced by \ttco10, spans 7\,pc in radius and suggests an outflow driven by a disk wind due to its asymmetrical structure along the minor axis. Collimated molecular outflows (CMO), traced by the low-velocity components of the \hcn32\ and \hcop32\ lines, align with the minor axis gas disk. Examination of position-velocity plots of \hcn32\ and \hcop32\ reveals a flared dense gas disk extended a radius of $\sim$\,60\,pc, infalling and rotating at speeds of about 200\,\kms\ and 300\,\kms\, respectively. A centrifugal barrier, located around 4\,pc from the dynamical center, implies an SMBH mass of approximately 10$^8$\,\msun, consistent with millimeter corona emission estimates. The CMO maintains a steady rotation speed of 200\,\kms\ over the 100\,pc scale along the minor axis. The projected speed of the CMO is about 80\,\kms, corresponding to around $\sim$\,500\,\kms, assuming an inclination angle of 80$^{\circ}$. Such a kinematics structure of disk-driven collimated rotating molecular outflow with gas supplies from a falling rotating disk indicates that the feedback of the compact obscured nucleus is likely regulated by the momentum transfer of the molecular gas that connects to both the feeding of the nuclear starburst and supermassive black hole.
%% add here a maximum of 10 keywords, to be taken form the file <Keywords.txt>
\end{abstract}

\firstsection % if your document starts with a section,
              % remove some space above using this command.
\section{Introduction}
Compact obscured nuclei (CONs) represent an extreme phase of galaxy evolution where rapid supermassive black hole growth and/or compact star-forming activity is completely obscured by gas and dust \citep[e.g.,][]{2013ApJ...764...42S, 2015A&A...584A..42A, 2019ApJ...882..153G, 2021A&A...649A.105F}. CONs are found in local luminous and ultraluminous infrared galaxies (U/LIRGs), with an occurrence rate of 38\% in ULIRGs and 21\% in LIRGs \citep{2021A&A...649A.105F}. These CONs are extremely dusty with high column densities that obscure X-ray to far-infrared emission and sometimes only leave optically thin windows at the submillimeter to radio bands \citep{2017ApJ...836...66S}. Rotational transitions of the vibrational excited HCN thus become a good tracer of CONs, which can probe deeper into the obscured nuclei, absorbing mid-infrared photons (e.g., 14\,$\mu$m, \citealt{2015A&A...584A..42A}) and radiate at submillimeter bands. CONs are, therefore, selected based on the HCN-vib lines, considering both their luminosity and surface density. The intense infrared radiation emanating from warm dust in CONs can account for a significant fraction of the bolometric luminosity of the galaxies. This radiation is thought to be powered either by the central active galactic nuclei (AGN) and/or an extreme nuclear starburst. The exact power source of CONs is not yet fully understood \citep{2021A&A...649A.105F}. Recent observations suggest that collimated molecular outflows (CMO) are ubiquitous in CONs \citep{2018ApJ...853L..28B, 2020A&A...640A.104A, 2023A&A...670A..70G}. However, the origin of the CMOs remains unclear. Understanding this could potentially shed light on the interplay between the AGN and nuclear starburst, as well as the correlation between the growth of the stellar nuclear mass and the mass of supermassive black holes (SMBHs).

To address these issues, we targeted IRAS\,17578-0400 ($L_\mathrm{IR}$\,=\,$2.3 \times 10^{11}$\,\lsun, redshift $z=0.0134$ with 284\,pc/\arcsec), one of the LIRGs identified as a CON from the CONquest survey \citep{2021A&A...649A.105F}. Using high-angular resolution ALMA Band-3 and Band-6 images, we are able to probe the spatial and kinematic structure of the ISM at scales of 5--12\,pc for the first time, providing important clues about the nature of CONs.
 
\vspace{-0.5cm}
\section{Structure of the CON in IRAS\,17578-0400 revealed by ALMA}

We have obtained high-angular resolution ALMA observations at Band 3 ($\sim$\,105\,GHz) and Band 6 ($\sim$\,255\,GHz), with angular resolutions reaching 0.06-0.03\arcsec\ and 0.04-0.02\arcsec (from natural weighting to uniform weighting), respectively.

\begin{figure}[hbtp!]
\vspace*{-0.35 cm}
\begin{center}
 \includegraphics[width=9cm]{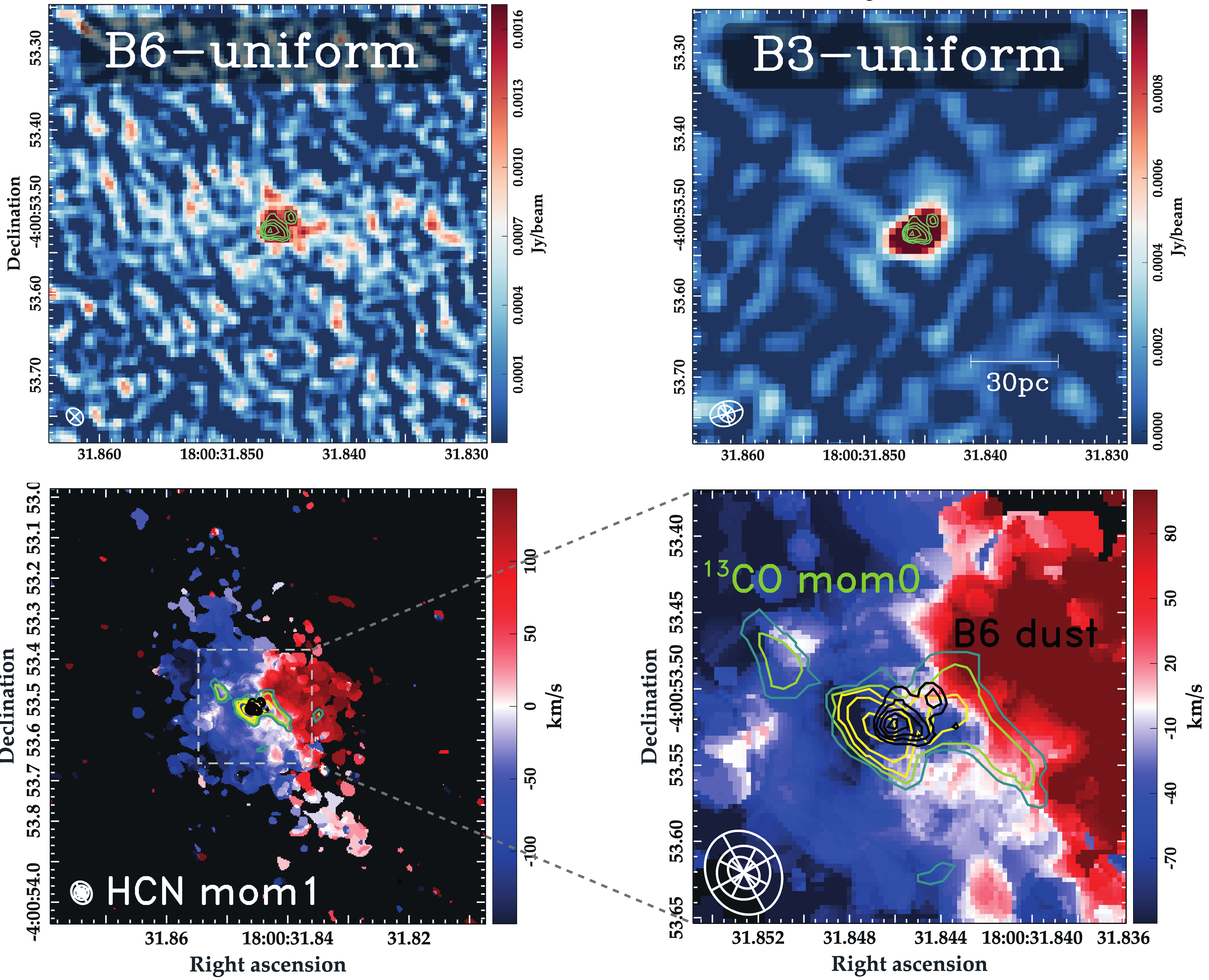}  
\vspace*{-0.35 cm}
 \caption{{\it Upper panels:} dust continuum in Band 6 and Band 3 from ALMA, imaged with uniform weighting to achieve the best resolution, to 0.02 and 0.03\arcsec, respectively. {\it Lower panels:} the moment-1 map of HCN(3--2), overlaid with yellow-green contours of the moment-0 map of $^{13}$CO and the black contours of the ALMA Band-6 dust continuum.}
   \label{fig1}
\end{center}
\end{figure}

{\underline{\it Structure of the ALMA continuum}}.  

As shown in the {\it upper panel} of Fig.\ref{fig1}, we find a very compact continuum structure from ALMA Band-3 and Band-6 images at scales of 20--30\,pc. Using uniform weighting, we are able to resolve the most compact structures in the continuum in Band-6, where the source breaks into two components, with a prominent east component and a western one. These two components are also seen by the elongated structure seen in the continuum image at Band-3. Such a structure suggests either a presence of a $\sim$\,4\,pc dusty torus around the AGN or a dual-AGN with a separation of $\sim$\,8\,pc. 

When examining the spectral energy distribution of the continuum, combined with the VJLA Ka and Ku band high-angular resolution images \citep{2022ApJ...940...52S}, we find the SED of the CON cannot be explained by a combination of the synchrotron, free-free, and the optically thick/thin dust. The flat ALMA Band-3 in-band spectral index ($\sim$\,1.2) and the ratio between ALMA Band-3 and Band-6 ($\sim$\,3) indicate an additional contribution from the millimeter corona emission \citep{2015MNRAS.451..517B} is significant. Our model, derived from the fitting of the corona emission, suggests an SMBH mass of $9.1^{+2.1}_{-3.8}\times10^{7}$\,\msun.

{\underline{\it Structure of the gas components}}. 

We highlight here the spatial structures of HCN(3--2), $^{13}$CO(1--0) on top of the dust continuum in the {\it lower panel} of Fig.\ref{fig1}. As in other CONs, HCN(3--2) shows a combination of a $\sim$\,60\,pc flared rotating disk with a $\sim$\,120\,pc CMO (north-south direction) extended perpendicular to the disk plane on each side. It is evident that $^{13}$CO shows a disk structure surrounding the dust continuum, and the velocity structure of $^{13}$CO suggests that it is a rotating disk with a radius of $\sim$\,7\,pc while the kinematic center is located in between the two dust components, supporting the scenario of a gas disk surrounding the dusty torus, rotating around the central SMBH (or a revolving binary SMBHs).

Besides the rotating disk, we also find tail-like asymmetrical outflow structures, with a blue-shifted "tail" in the northeast and a red-shifted tail towards the southwest (Fig.\ref{fig1}). Similar structures have been found in Galactic circumstellar disks \citep{2017A&A...603L...3A}. Such an asymmetry is caused by a disk-launching wind with infalling material onto the disk asymmetrically. Considering the $^{13}$CO wind is connected to the HCN(3--2) CMO, the structure of $^{13}$CO indicates that the CMO is likely also driven by the rotating disk.

\begin{figure}[hbtp]
\vspace*{-0.35 cm}
\begin{center}
 \includegraphics[width=11cm]{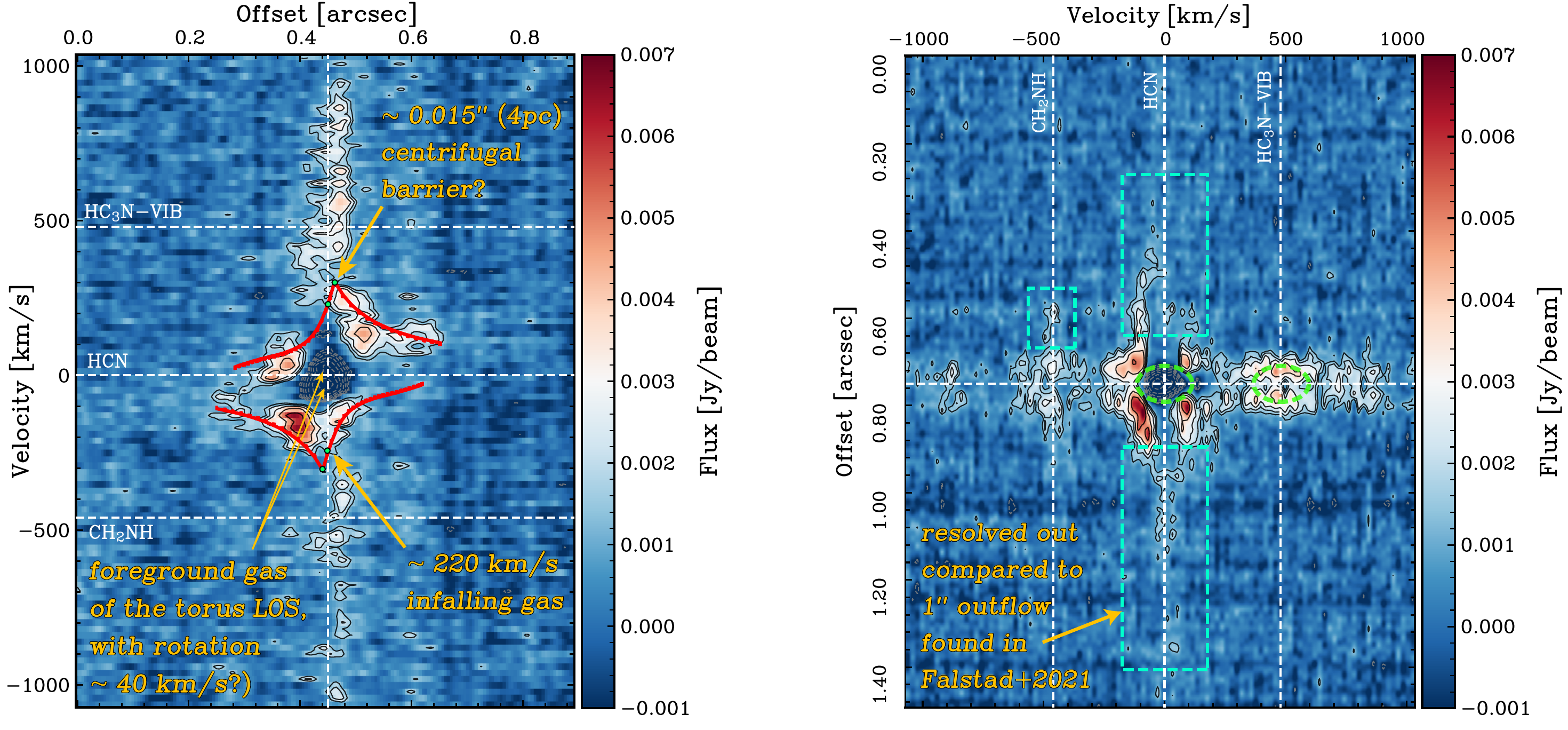} 
\vspace*{-0.35 cm}
 \caption{PV plots of HCN(3--2) of IRAS\,271578-0400. The Left and right panels are extracted from the major and minor axes, respectively. We also indicate the line centers using white dashed lines. The red lines indicate a model of an infalling rotating disk.}
   \label{fig2}
\end{center}
\end{figure}

{\underline{\it Kinematics}}. 

To further understand the kinematics of the CON and CMO, we have examined the structure of the position-velocity (PV) plots of HCN(3--2), as shown in Fig.\,\ref{fig2}. The major-axis PV plot reveals bright emissions from non-Keplerian motions. We find that an infalling rotating disk model \citep{2014Natur.507...78S} can well explain the PV structure. From the model, we derive an overall rotating speed of the flared gas disk of $\sim$\,300\,\kms\ with an infalling velocity of $\sim$\,200\,\kms, a centrifugal barrier radius of $\sim$\,4\,pc (where the rotation speed is $\sim$\,300\,\kms), suggesting an SMBH mass of $10^{8}$\,\msun, consistent with the estimate from millimeter corona emission. From the minor-axis PV plot, we also find that the CMO extends to about 120\,pc to the minor-axis direction, which is more than half smaller than the CMO identified with a lower-spatial resolution data extending to $\sim$\,300\,pc \citep{2021A&A...649A.105F}, possibly because the emission is resolved out. Offset (w.r.t the major axis) PV cuts perpendicular to the CMO direction also reveal that the CMO is rotating at the speed of $\sim$\,200\,pc with little slowing down.

\section{Conclusions and implications}

\begin{figure}[hbtp]
\vspace*{-0.35 cm}
\begin{center}
 \includegraphics[width=5.2cm]{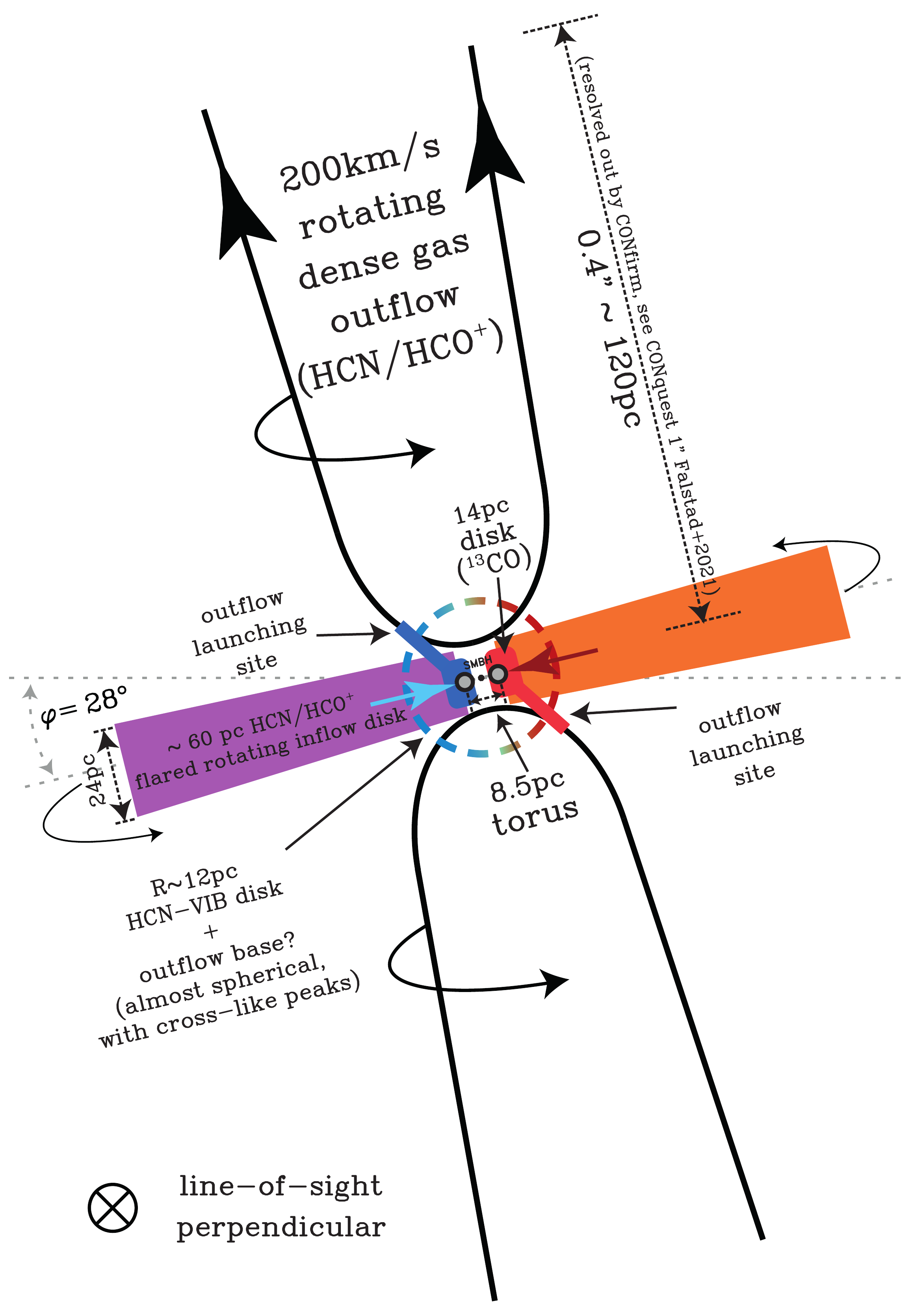} 
\vspace*{-0.35 cm}
 \caption{A cartoon of the structure of the CON in IRAS\,17578-0400.}
   \label{fig3}
\end{center}
\end{figure}

Using high-spatial-resolution ALMA data, we study the spatial and kinematic structure of the compact obscured galactic nucleus of IRAS\,17578-0400. We are able to resolve down to a few parsecs for the first time. Combining all the spatial and kinematic information together, we present a coherent model in Fig.\,\ref{fig3} of the nature of the CON. We identify the key kinematic structure is dominated by a flared rotating infalling disk about 60\,pc in size from the outer region towards the inner 4\,pc where it reaches the centrifugal barrier, and at this radius, we also see the dust torus from the continuum image. Such a rotating disk is also likely driving the CMO that launched to both sizes of the minor axis. Such a structure might be a key driver of feedback processes, transferring the gas into the central region and feeding the nuclear starburst and SMBH, linking the growth of both.

\vspace{-0.5cm}
\acknowledgments
The authors acknowledge the support from the ERC Advanced Grant 789410.

\end{document}